\documentclass[12pt,preprint]{aastex}
\def\msun{{\rm M_{\odot}}}

\def\be{\begin{equation}}
\def\ee{\end{equation}}

\def\le{{L_{\rm Edd}}}
\def\msun{{\rm M_{\odot}}}

\def\mo{{\dot M_{\rm out}}}
\def\me{{\dot M_{\rm Edd}}}
\def\tc{{t_{\rm C}}}
\begin{document}

\title{BLACK HOLES, GALAXY FORMATION, AND THE $M_{\rm BH} - \sigma$
 RELATION}

\author{ Andrew~King\altaffilmark{1, 2}}

\altaffiltext{1} {Department of Physics and Astronomy, University of
Leicester, Leicester LE1 7RH, U.K.; ark@astro.le.ac.uk}
\altaffiltext{2} {Harvard--Smithsonian Center for Astrophysics, 60
Garden Street, Cambridge, MA 02138}

\begin{abstract}

Recent X--ray observations of intense high--speed outflows in quasars
suggest that supercritical accretion on to the central black hole may
have an important effect on a host galaxy. I revisit some ideas of
Silk and Rees, and assume such flows occur in the final stages of
building up the black hole mass. It is now possible to model
explicitly the interaction between the outflow and the host
galaxy. This is found to resemble a momentum--driven stellar wind
bubble, implying a relation $M_{\rm BH} = (f_g\kappa/2\pi G^2)\sigma^4
\simeq 1.5\times 10^8\sigma_{200}^4~\msun$ between black hole mass and
bulge velocity dispersion ($f_g =$ gas fraction of total matter
density, $\kappa =$ electron scattering opacity), without free
parameters. This is remarkably close to the observed relation in both
slope and normalization. 

This result suggests that the central black holes in galaxies gain
most of their mass in phases of super--Eddington accretion, which are
presumably obscured or at high redshift. Observed super--Eddington
quasars are apparently late in growing their black hole masses.

\end{abstract}


\section{Introduction}
It is now widely accepted that the centre of every galaxy contains a
supermassive black hole. The close observational correlation 
(Ferrarese \& Merritt, 2000; Gebhardt et al., 2000; Tremaine et al., 2002)
between the mass $M$ of this hole and the velocity dispersion $\sigma$ of 
the host bulge strongly suggests a connection between the formation of
the black hole and of the galaxy itself.

Recent {\it XMM--Newton} observations of bright quasars (Pounds et
al., 2003a, b; Reeves et al., 2003) may offer a clue to this
connection. These observations give strong evidence for intense
outflows from the nucleus, with mass rates $\mo \sim 1\msun~{\rm
yr}^{-1}$ and velocity $v \sim 0.1c$, in the form of blueshifted
X--ray absorption lines. Simple theory shows that the outflows are
probably optically thick to electron scattering, with a photosphere of
$\sim 100$ Schwarzschild radii, and driven by
continuum radiation pressure. In all cases the outflow velocity is
close to the escape velocity from the scattering photosphere. As a
result the outflow momentum flux is comparable
to that in the Eddington--limited radiation field, i.e.
\begin{equation}
\mo v \simeq {\le\over c},
\label{mom}
\end{equation}
where $\mo$ is the mass outflow rate and $\le$ the Eddington
luminosity, while the mechanical energy flux is
\begin{equation}
{1\over 2}\mo v^2 \simeq {\le^2\over 2\mo c^2}. 
\label{en}
\end{equation}

It appears that such outflows are a characteristic of super--Eddington
accretion (King \& Pounds, 2003). We know that most of the mass of the
nuclear black holes is assembled by luminous accretion (Soltan 1982;
Yu \& Tremaine, 2002). It seems likely that the rate at which mass
tries to flow in towards the central black hole in a galaxy is set by
conditions far from the hole, for example by interactions or mergers
with other galaxies. It is quite possible therefore that
super--Eddington conditions prevail for most of the time that the
central black hole mass is being built up.

This clearly has important implications for the
host galaxy. Unlike luminous energy, a large fraction of a mechanical
energy flux like
(\ref{en}) is likely to be absorbed within the galaxy, and must have a
major effect. To reach its present mass the black hole in
PG1211+143 could have accreted at a rate comparable to its
current one for $\sim 5\times 10^7$~yr. During that time, an outflow
like the observed one could have deposited almost
$10^{60}$~erg in the host galaxy. This exceeds the binding energy
$\sim 10^{59}$~erg of a bulge with $10^{11}~\msun$ and $\sigma 
\sim 300~{\rm km~s}^{-1}$.  

Accordingly it is appropriate to revisit some ideas presented by Silk
and Rees (1998., henceforth SR98) and also considered by Haehnelt et
al. (1998), Blandford (1999) and Fabian (1999).
These authors envisage a situation in which
the initial black
holes formed with masses $\sim 10^6\msun$ before most of the stars.
Accretion on to these black holes is assumed to produce outflow, which
interacts with the surrounding gas.
Without a detailed treatment of the outflow from a
supercritically accreting black hole, SR98 used dimensional arguments
to suggest a relation between $M$ and $\sigma$. However this still has
a free parameter. Given the simple relation
(\ref{mom}) one can now remove this freedom. The situation turns out
to resemble a momentum--driven stellar wind bubble. Modelling this
gives an $M_{\rm BH} - \sigma$ relation devoid of free parameters, and 
remarkably close to the observed relation.

\section{Black Hole Wind Bubbles}

I follow SR98 in modelling a protogalaxy as an isothermal sphere of
dark matter. If the gas fraction is $f_g = \Omega_{\rm
baryon}/\Omega_{\rm matter}\simeq 0.16$ (Spergel et al., 2003) its
density is
\begin{equation}
\rho = {f_g\sigma^2\over 2\pi Gr^2}
\label{rho}
\end{equation}
where $\sigma$ is assumed constant. The gas mass inside radius $R$ is
\begin{equation}
M(R) = 4\pi\int_0^R\rho r^2 {\rm d}r = {2f_g\sigma^2R\over G}
\label{m}
\end{equation}
I assume that mass flows towards the central black
hole at some supercritical rate $\dot M_{\rm acc}$. The
results of King \& Pounds (2003) suggest that this will produce a
quasi--spherical outflow with momentum flux given by (\ref{mom}). Note
that this momentum rate is independent of the outflow rate
$\mo = \dot M_{\rm acc} - \me$ since
the outflow velocity $v$ adjusts as $\mo^{-1}$ to maintain the
relation (\ref{mom}) (King \& Pounds, 2003). 

The wind from the central black hole will sweep up the surrounding gas
into a shell. As is well known from the theory of stellar wind bubbles
(e.g. Lamers \& Casinelli 1999) this shell is bounded by an inner shock where
the wind velocity is thermalized, and an outer shock where the surrounding
gas is heated and compressed by the wind. These two regions are
separated by a contact discontinuity. The shell velocity depends on whether
the shocked wind gas is able to cool (`momentum--driven' flow) or not
(`energy--driven' flow). In the absence of a detailed treatment of a
quasar wind, SR98 appear to have assumed the second case. In fact for the
supercritical outflows envisaged here the 
first case is more likely, as the argument below shows.

\section{Cooling the Wind Shock}

The Compton cooling time of an electron of energy $E$ is
\begin{equation}
\tc = {3m_ec\over 8\pi\sigma_{\rm T}U_{\rm rad}}{m_ec^2\over E}
\label{compt}
\end{equation}
where $m_e$ is the electron mass and
\begin{equation}
U_{\rm rad} = {\le\over 4\pi R^2cb}
\label{urad}
\end{equation}
is the radiation density at distance $R$ from the black hole, and $b
\la 1$ allows for some collimation of the outflow. The
electron energy $E$ in the postshock wind gas is $\simeq 9m_pv^2/16$, where
$v$ is the wind velocity and $m_p$ the proton mass. 
Combining this with the usual definition
\begin{equation} 
\le = {4\pi GM_{\rm BH}c\over \kappa}
\label{edd}
\end{equation}
of the Eddington luminosity for black hole mass $M_{\rm BH}$ shows that 
\begin{equation}
\tc = {2\over 3}{cR^2\over GM}\biggl({m_e\over m_p}
\biggr)^2\biggl({c\over v}\biggr)^2b
\simeq 10^5R_{\rm kpc}^2\biggl({c\over v}\biggr)^2bM_8^{-1}~{\rm yr}
\label{tc}
\end{equation}
where $R_{\rm kpc}$ is $R$ measured in kpc and $M_8 =
M_{\rm BH}/10^8\msun$. Clearly this is extremely short for small $R$,
so the flow is efficiently cooled and thus momentum driven at least
initially. I note that Ciotti \& Ostriker (1997, 2001) emphasize
the importance of Compton heating and cooling on quasar inflows and outflows.

The momentum--driven assumption breaks down once $\tc$ becomes of order the
flow time $t_{\rm flow} = R/v_s$ where $v_s$ is the shell velocity. 
We can use the momentum--driven shell velocity $v_m$ derived in eqn
(\ref{vs}) below to estimate
\begin{equation}
t_{\rm flow} = 8\times 10^6R_{\rm kpc}\sigma_{200}M_8^{-1/2}~{\rm yr}
\end{equation}
where $\sigma_{200} = \sigma/(200~{\rm km~s}^{-1})$. 
The assumption of efficient cooling is valid out to a radius $R_c$ given
by setting $\tc = t_{\rm flow} = 1$. We find a total 
swept--up mass
\begin{equation}
M(R_c) = 1.9\times 10^{11}\sigma_{200}^3M_8^{1/2}
\biggl({v\over c}\biggr)^2b^{-1}~\msun
\label{mr}
\end{equation}
at this point. Once the shell reaches radii larger than $R_c$ the
shocked wind is no longer efficiently cooled, and its thermal pressure
accelerates the shell of swept--up gas to a higher velocity $v_e > v_m$ 
(energy--driven flow) after a sound--crossing time $\sim R_c/v$.

\section{The $M_{\rm BH} - \sigma$ relation}

I now estimate the speed $v_m$ of the momentum--driven shell by the
standard wind bubble argument. At sufficiently large radii $R$ the
swept--up shell mass $M(R)$ is much larger than the wind mass, and the
shell expands under the impinging wind ram pressure $\rho v^2$ (this
characterizes momentum--driven flows; in an energy--driven flow the
thermal pressure of the shocked wind gas is dominant, while in a
supernova blast wave the momentum injection is instantaneous rather
than continuous). The shell's equation of motion is thus
\begin{equation}
{{\rm d}\over {\rm d}t}\biggl[M(R)\dot R\biggr] = 
4\pi R^2\rho v^2 = \mo v = {\le\over c}
\label{motion}
\end{equation}
where we have used first the mass conservation equation for the quasar
wind, and then (\ref{mom}) to simplify the rhs. Integrating this
equation for $\dot R$ with the final form of the rhs gives
\begin{equation}
M(R)\dot R = {\le\over c}t
\end{equation}
where I have neglected the integration constant as $M(R)$ is
dominated by swept--up mass at large $t$. Using (\ref{m})
for $M(R)$ and integrating
once more gives
\begin{equation}
R^2 = {G\le \over 2f_g\sigma^2c}t^2,
\end{equation}
where again we may neglect the integration constant for large
$t$. We see that in the snowplow phase the shell moves with constant
velocity $v_m = R/t$, with
\begin{equation}
v_m^2 = {G\le \over 2f_g\sigma^2c}.
\label{vs}
\end{equation}

We note that this velocity is larger for higher $\le$, i.e. higher
black hole mass. This solution holds if the shell is inside the
cooling radius $R_c$; outside this radius the shell speed eventually
increases to the energy--driven value $v_e$, which also grows with
$M_{\rm BH}$.

I now consider the growth of the black hole mass by accretion.
Initially the mass is small, inflow is definitely supercritical, and
even the energy--driven shell velocity would be smaller than the
escape velocity $\sigma$. No mass is driven away, and accretion at a
rate $\dot M_{\rm Edd}$ can occur efficiently.  However as the black
hole grows, we eventually reach a situation in which $v_e > \sigma >
v_m$. Further growth is now only possible until the shell
reaches $R_c$, and then only until the point where $v_m =
\sigma$. Thus given an adequate mass supply, e.g. through
mergers, the final black hole mass is given by setting $v_m =
\sigma$ in (\ref{vs}). Using (\ref{edd}) we find the relation
\begin{equation}
M_{\rm BH} = {f_g\over 2\pi}{\kappa\over G^2}\sigma^4 \simeq
1.5\times 10^8\sigma_{200}^4~\msun. 
\label{msigma}
\end{equation}
This is remarkably close to the observed relation (Tremaine et al.,
2000). 

Presumably most of the swept--up mass ends up as bulge stars, and we
may tentatively identify $M(R_c)$ as an upper limit the bulge mass
$M_b$ of the galaxy. Using (\ref{msigma}) to eliminate $\sigma_{200}$
we get
\begin{equation}
M_{\rm BH} \ga 7\times 10^{-4}M_8^{-1/4}\biggl({c\over v}\biggr)^2bM_b.
\label{mg}
\end{equation}
If $c/v$ (determined by the ratio $\dot M_{\rm out}/\dot M_{\rm Edd}$)
attains similar values at this point in most systems and the swept--up
mass is close to $M(R_c)$ one gets a relation between black hole and
bulge mass of the form $M_b \propto M_{\rm BH}^{1.25}$.  The relation
is written instead in the form (\ref{mg}) to allow easy comparison
with the correlation found by Magorrian et al.  (1998). Evidently this
is not as clear--cut a relation as that between $M_{\rm BH}$ and
$\sigma$, and indeed the scatter in the observed relation is
considerably larger.

\section{Discussion}

The $M_{\rm BH} - \sigma$ relation given here has no free
parameter. If the outflow velocity $v$ had been larger by an optical
depth factor $\tau> 1$ (i.e. most of the acceleration occurs below the
photosphere) a factor $1/\tau$ would appear on the rhs. However this
would require outflow velocities $\tau (GM/R_{\rm ph})^{1/2}$ larger
by the same factor than those observed in supercritically accreting
quasars.

The lack of freedom in (\ref{msigma}) comes about because the physical
situation envisaged by SR98 and also studied by Haehnelt et al., 1998)
and Blandford (1999) can now be made more precise: the response of
observed black hole systems to super--Eddington accretion appears to be
an optically thick outflow driven by continuum radiation
pressure. Fabian (1999) considers {\it sub}--Eddington accretion, but
emphasizes the importance of the momentum of the outflow as opposed to
its energy: it is this which leads to the $\sigma^4$ dependence rather
than $\sigma^5$. Specifically, Fabian (1999) assumes a wind of speed
$v_w$ with mechanical luminosity a fixed fraction $a$ of $\le$. This
produces a relation of the form (\ref{msigma}) but with an extra
factor $v_w/ac$ on the rhs; it therefore reduces to (\ref{msigma}) if
one assumes $a \sim v_w/c$.  Parameters also appear in other
derivations using different physics, such as the ambient conditions in
the host galaxy (Adams, Graff \& Richstone, 2001) or accretion of
collisional dark matter (Ostriker, 2000).

The picture presented here invokes a largely spherical geometry for
the ambient gas, except that the accreting matter must possess a small
amount of angular momentum to define an accretion disc plane and thus
a small solid angle where inflow rather than outflow occurs. It is
therefore appropriate to the growth of a spheroid--black hole
system. However once most of the gas lies in the plane of the galaxy
the momentum--driven outflow considered here would not halt
inflow. Evidently this means that accretion from this point on adds
little mass to the hole.

If the derivation of the $M_{\rm BH} - \sigma$ relation given here is
some approximation to reality, it implies that the central black holes
in galaxies gain most of their mass in phases of super--Eddington
inflow. As relatively few AGN are observed in such phases, these must
either be obscured (cf Fabian, 1999) or at high redshift. It appears
then that those quasars which are apparently now accreting at such
rates (Pounds et al., 2003a,b; Reeves et al., 2003) are laggards in
gaining mass. This idea agrees with the general picture that these
objects -- all narrow--line quasars -- are super--Eddington because
they have low black--hole masses, rather than unusually high mass
inflow rates.

\acknowledgements

I thank Andy Fabian, Jim Pringle and James Binney for illuminating
discussions, and the referee for a very helpful report. This work was
carried out at the Center for Astrophysics, and I thank members of its
staff, particularly Pepi Fabbiano and Martin Elvis, for stimulating
discussions and warm hospitality.  I gratefully acknowledge a Royal
Society Wolfson Research Merit Award.

\end{document}